\documentclass[letterpaper,english,preprint, aps]{revtex4-1}
\usepackage[T1]{fontenc}
\usepackage[latin9]{inputenc}
\usepackage{mathrsfs}
\usepackage{amsmath}
\usepackage{amssymb}
\usepackage{graphicx}

\makeatletter

\pdfpageheight\paperheight
\pdfpagewidth\paperwidth

\makeatother

\usepackage{babel}
\begin{document}

\preprint{BROWN-HET-1677}

\title{Black hole mining in the RST model}

\author{Rohitvarma Basavaraju}
\email{rohitvarma_basavaraju@brown.edu}

\author{David A. Lowe}
\email{lowe@brown.edu}

\affiliation{Department of Physics, Brown University, Providence, RI 02912, USA}

\date{\today}
\begin{abstract}
We consider the possibility of mining black holes in the 1+1-dimensional
dilaton gravity model of Russo, Susskind and Thorlacius. The model
correctly incorporates Hawking radiation and back-reaction in a semiclassical
expansion in $1/N$, where $N$ is the number of matter species. It
is shown that the lifetime of a perturbed black hole is independent
of the addition of any extra apparatus when realized by an arbitrary
positive energy matter source. We conclude that mining does not occur
in the RST model and comment on the implications of this for the black
hole information paradox.
\end{abstract}


\maketitle

\section{Introduction}

Hawking showed that black holes radiate quantum mechanically and eventually
evaporate \citep{Hawking:1974sw}. It is interesting to wonder if
there is a consistent manner in which one might perturb a black hole
to cause it to evaporate more quickly. Unruh and Wald \citep{UnruhWald,UnruhWald2}
have argued that radiation from black holes can be mined rapidly,
potentially diminishing the lifetime of a 4-dimensional Schwarzschild
black hole of mass $M$ to a time of order $M$. The unperturbed lifetime
on the other hand is of order $M^{3}$. Such a drastic drop raises
thorny questions concerning the black hole information paradox, where
apparently one needs at least a scrambling time of order $M\log M$
before quantum information may be carried off by the Hawking radiation
\citep{Hayden:2007cs,Sekino:2008he,Lowe:2015eba,Lowe:2016mhi}. By
considering optimal semiclassical solutions that accomplish mining,
\citep{Brown:2012un} suggests that such a process may at most diminish
the lifetime of a black hole to of order $M^{2}$, safely greater
than the scrambling time, but parametrically shorter than the unperturbed
lifetime.

In the present work our goal is not to test the idea of rapid mining,
but simply to consider whether any mining is possible in a model where
the question is fully tractable. The original model computation of
mining presented in \citep{UnruhWald2} was motivated by a two-dimensional
moving mirror in flat spacetime. The models of two-dimensional dilaton
gravity provide an ideal testing ground for this particular question,
which are a natural generalization of the moving mirror model. In
particular, the RST model \citep{Russo:1992yh} incorporates a reflecting
boundary condition, as well as a full treatment of back-reaction of
the Hawking radiation on the geometry at the semiclassical level.
Using the scalar matter fields of RST, our strategy will be to send
an arbitrary scalar waveform into an evaporating black hole spacetime.
We then check if the black hole lifetime decreases. Although there
exists a literature on energy conservation in RST \citep{Kim:1995jta,Kim:1995wr},
the connection with mining and the computation of the lifetime for
general flux is new.

\section{Review of Semiclassical Two-Dimensional Dilaton Gravity Coupled to
Matter in RST Model}

As our basic approach is to consider the effect of arbitrary flux
on a black hole in the RST model, we review the relevant elements
of that model here. A useful general review of these models is \citep{Strominger:1994tn}.

\subsection{Action and Equations of Motion}

The RST action can be thought of as a one loop quantum corrected version
of the dilaton gravity action due to Callan, Giddings, Harvey and
Strominger (CGHS)\citep{Callan:1992rs}:

\begin{align*}
S & =\frac{1}{2\pi}\int d^{2}x\sqrt{-g}[e^{-2\phi}(R+4(\nabla\phi)^{2}+4\lambda^{2})-\frac{1}{2}\sum_{i=1}^{N}(\nabla f_{i})^{2}]\\
 & -\frac{\kappa}{8\pi}\int d^{2}x\sqrt{-g}[R\frac{1}{\nabla^{2}}R+2\phi R].
\end{align*}
The action for the RST model is given by:

\begin{align}
S & =\frac{1}{\pi}\int d^{2}x[e^{-2\phi}(2\partial_{+}\partial_{-}\rho-4\partial_{+}\phi\partial_{-}\phi+\lambda^{2}e^{2\rho})\nonumber \\
 & +\frac{1}{2}\sum_{i=1}^{N}\partial_{+}f_{i}\partial_{-}f_{i}-\frac{\kappa}{\pi}(\partial_{+}\rho\partial_{-}\rho+\phi\partial_{+}\partial_{-}\rho)]\label{eq:action}
\end{align}
in conformal gauge where the only non-vanishing components of the
metric are $g_{-+}=g_{+-}=-\frac{1}{2}e^{2\rho}.$ $\phi$ is the
dilaton and $f_{i}$ are matter scalar fields. The constants $\kappa=\frac{N\hbar}{12\pi}$
and $\lambda$ play the roles of Planck's constant and the cosmological
constant in this model. We will generally set the latter to unity
for convenience. We work in null Kruskal-type coordinates. The $\kappa$
dependent terms in \eqref{eq:action} represent the effect of integrating
out the one-loop fluctuations of matter fields, as well as the RST
improvement term, responsible for maintaining the solubility of the
equations of motion. We note the constraint equations in this gauge
couple the matter fields to the gravitational fields $\rho$ and $\phi$

\begin{align*}
0 & =T_{\pm\pm}=(e^{-2\phi}-\frac{\kappa}{4})(4\partial_{\pm}\rho\partial_{\pm}\phi-2\partial_{\pm}^{2}\phi)+\frac{1}{2}\sum_{i=1}^{N}\partial_{\pm}f_{i}\partial_{\pm}f_{i}\\
 & -\kappa(\partial_{\pm}\rho\partial_{\pm}\rho-\partial_{\pm}^{2}\rho+t_{\pm}).
\end{align*}
The functions $t_{\pm}$ are fixed by boundary conditions on the stress
energy tensor. We identify the matter contribution to the stress tensor
as:

\[
T_{\pm\pm}^{f}=\frac{1}{2}\sum_{i=1}^{N}\partial_{\pm}f_{i}\partial_{\pm}f_{i}.
\]

The following field redefinitions lead to especially simple equations
of motion:

\[
\Omega=\frac{\sqrt{\kappa}}{2}\phi+\frac{e^{-2\phi}}{\sqrt{\kappa}}
\]

\[
\chi=\sqrt{\kappa}\rho-\frac{\sqrt{\kappa}}{2}\phi+\frac{e^{-2\phi}}{\sqrt{\kappa}}\,.
\]

The resulting action, equations of motion and constraints are:

\[
S=\frac{1}{\pi}\int d^{2}x\left[-\partial_{+}\chi\partial_{-}\chi+\partial_{+}\Omega\partial_{-}\Omega+\lambda^{2}e^{\frac{2}{\sqrt{\kappa}}(\chi-\Omega)}+\frac{1}{2}\sum_{i=1}^{N}\partial_{+}f_{i}\partial_{-}f_{i}\right]
\]

\[
\partial_{+}\partial_{-}\chi=\partial_{+}\partial_{-}\Omega=-\frac{\lambda^{2}}{\sqrt{\kappa}}e^{\frac{2}{\sqrt{\kappa}}(\chi-\Omega)}
\]

\begin{equation}
\kappa t_{\pm}=-\partial_{\pm}\chi\partial_{\pm}\chi+\sqrt{\kappa}\partial_{\pm}^{2}\chi+\partial_{\pm}\Omega\partial_{\pm}\Omega+\frac{1}{2}\sum_{i=1}^{N}\partial_{\pm}f_{i}\partial_{\pm}f_{i}.\label{eq:constraints}
\end{equation}
We further fix the gauge to Kruskal gauge where $\chi=\Omega.$ 

\subsection{On Boundaries, Singularities and the Apparent Horizon}

In calculating the effect of flux on the black hole lifetime in RST,
we will be making use of the boundary of the spacetime, the linear
dilaton vacuum (LDV), and the positions of the black hole singularity
and apparent horizon. We now review these concepts and their relations. 

The LDV solution is the analog of higher dimensional flat spacetime
and is given by

\begin{equation}
\Omega=\frac{x^{+}x^{-}}{\sqrt{\kappa}}-\frac{\sqrt{\kappa}}{4}\log(-x^{+}x^{-}).\label{eq:ldv}
\end{equation}
The scalar curvature is given by:

\begin{equation}
R=\frac{8[1-(\nabla\phi)^{2}]}{\sqrt{\kappa}e^{2\phi}\Omega'(\phi)}.\label{eq:scalarcurv}
\end{equation}
This will be generically singular at the critical value $\phi=\phi_{cr}$
where $d\Omega/d\phi=0$ which we denote by the curve $\hat{x}^{+}(x^{-})$.

The apparent horizon is the boundary of a trapped region. The variable
$e^{-\phi}$ can be thought of as a analog of a transverse sphere
in the dimensional reduction of four-dimensional Einstein gravity
to two-dimensional dilaton gravity. The quantity $\nabla\phi$ will
therefore be timelike in a trapped region. The boundary is determined
by the equation $\left(\nabla\phi\right)^{2}=0$ which becomes
\[
\partial_{+}\Omega=0.
\]

In both subcritical and supercritical regimes, there is a unique curve
$\phi=\phi_{cr}$. In the former case, this is the boundary of the
LDV. In the latter case, it is the black hole singularity where \eqref{eq:scalarcurv}
diverges. To maintain cosmic censorship, RST observe that finite curvature
at the boundary requires imposing the cosmic censorship boundary condition

\begin{equation}
\partial_{\pm}\Omega=0\,.\label{eq:rstbc}
\end{equation}
This implies reflective boundary conditions for the energy flux 

\begin{equation}
T_{--}^{f}-\kappa t_{-}=\left(\frac{d\widehat{x}^{+}}{dx^{-}}\right)^{2}\left(T_{++}^{f}-\kappa t_{+}\right).\label{eq:bcondition}
\end{equation}

\section{Generalizing RST to arbitrary matter flux}

We review here the reflection of subcritical flux and black hole formation
from supercritical flux. These cases constitute the general solutions
to RST for arbitrary matter flux in the strictly subcritical and supercritical
regimes. We then consider a general black hole solution with general
ingoing subcritical flux. 

\subsection{Review of Subcritical and Supercritical Cases}

\begin{figure}
\includegraphics[scale=0.8]{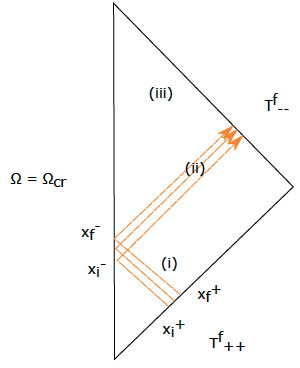}

\caption{\label{fig:Reflection-of-subcritical}Reflection of subcritical flux
in RST with three distinct regions corresponding to infall of flux,
reflection and LDV aftermath. }
\end{figure}

\begin{figure}
\includegraphics[scale=0.6]{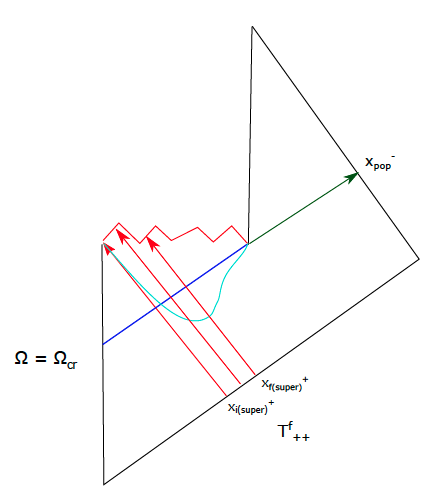}\caption{\label{fig:Black-hole-formation}Black hole formation by supercritical
flux, shown in red. Event horizon shown in blue, apparent horizon
in turquoise, thunderpop in green and black hole singularity in red. }
\end{figure}

A subcritical flux obeys the following inequality \citep{Russo:1992ax}

\[
T_{++}^{f}<\frac{\kappa}{4(x^{+})^{2}}\,.
\]
When the flux always obeys this condition the boundary remains timelike
and a black hole does not form. Such flux can be made to reflect from
the boundary of the spacetime given appropriate boundary conditions
as shown in figure \ref{fig:Reflection-of-subcritical}. The opposite
inequality will imply that the boundary curve becomes a spacelike
singularity, which is identified as the black hole singularity. We
refer to this latter case, where black hole formation occurs, as ``supercritical''
(see figure \ref{fig:Black-hole-formation}). 

For illustrative purposes, we review the manner in which purely subcritical
flux is reflected in RST. In the subsequent section the same procedure
will be used to reflect flux outside the event horizon of a black
hole. We will assume the ingoing subcritical flux is localized in
some region $x_{i}^{+}<x^{+}<x_{f}^{+}.$ The Bondi energy $M_{+}(x^{+})$
and the Kruskal momentum $P_{+}(x^{+})$ are defined as integrals
along $\mathscr{I}^{-}$

\[
P_{+}(x^{+})=\int_{x_{i}^{+}}^{x^{+}}T_{++}^{f}dx^{+}
\]

\[
M_{+}(x^{+})=\int_{x_{i}^{+}}^{x^{+}}T_{++}^{f}x^{+}dx^{+}.
\]
In fact, only these two integrals are needed since in our choice of
gauge the flux is related to the double derivative of $\Omega$:

\[
-\sqrt{\kappa}\partial_{\pm}^{2}\Omega=-\kappa t_{\pm}+\frac{1}{2}\sum_{i=1}^{N}\partial_{\pm}f_{i}\partial_{\pm}f_{i}.
\]

In the case of subcritical flux, one can trace out the reflection
of the flux from the boundary as shown in figure \ref{fig:Reflection-of-subcritical}.
There are three regions of interest separated by lines of constant
$x^{-}$, namely $x^{-}=x_{i}^{-}$ and $x^{-}=x_{f}^{-}$, defined
via $x_{i}^{+}=\hat{x}^{+}(x_{i}^{-})$ and $x_{f}^{+}=\hat{x}^{+}(x_{f}^{-})$.

The first region contains only infalling flux:

\[
\Omega^{(i)}(x^{-}<x_{i}^{-})=-\frac{x^{+}(x^{-}+P_{+}(x^{+}))}{\sqrt{\kappa}}-\frac{\sqrt{\kappa}}{4}\log(-x^{+}x^{-})+\frac{M_{+}(x^{+})}{\sqrt{\kappa}}.
\]
The function $t_{+}=\frac{1}{4x^{+2}}$ is determined by demanding
the solution approach the linear dilaton vacuum as $x^{-}\to-\infty$.
The second region corresponds to the reflection of flux, with \eqref{eq:bcondition}
determining the outgoing flux $T_{--}=T_{--}^{f}-\kappa t_{-}$

\begin{align*}
\Omega^{(ii)}(x_{i}^{-} & <x^{-}<x_{f}^{-})=-\frac{x^{+}(x^{-}+P_{+}(x^{+}))}{\sqrt{\kappa}}-\frac{\sqrt{\kappa}}{4}\log(-x^{+}x^{-})\\
 & +\frac{M_{+}(x^{+})}{\sqrt{\kappa}}+F(x^{-})
\end{align*}

\[
F(x^{-})=\frac{\sqrt{\kappa}}{4}\log\left(-x^{-}\hat{x}^{+}\right)-\frac{M\left(\hat{x}^{+}\right)}{\sqrt{\kappa}}-\frac{\sqrt{\kappa}}{4}\log(\frac{\kappa}{4}).
\]
The function $F(x^{-})$ and the shape of the boundary curve $\hat{x}^{+}(x^{-})$
is determined from the finite curvature conditions \eqref{eq:rstbc}.
The third region is the empty linear dilaton vacuum (where we denote
the constants $P_{+}^{f}\equiv P(x_{f}^{+})$ and $M_{+}^{f}\equiv M(x_{f}^{+}))$:

\[
\Omega^{(iii)}(x^{-}>x_{f}^{-})=-\frac{x^{+}(x^{-}+P_{+}^{f})}{\sqrt{\kappa}}-\frac{\sqrt{\kappa}}{4}\log(-x^{+}(x^{-}+P_{+}^{f})).
\]
Having reviewed the manner in which subcritical flux is reflected
with RST boundary conditions, as well as how black hole formation
occurs, we now turn to an analysis of a general black hole solution
\textit{with} additional subcritical flux probing the black hole.

\subsection{\label{subsec:General-Black-Hole}General Black Hole Solution and
Thunderpop}

The endpoint of Hawking radiation in RST results in a well known emission
of energy referred to as the ``thunderpop'' \citep{Russo:1992ax}.
We take the viewpoint that the thunderpop is to be viewed as a condition
to be imposed on the low energy theory to make it semiclassically
consistent, representing the effect of some higher derivative interactions
that descend from some unknown ultraviolet complete theory. This allows
us to avoid the issues raised in \citep{Strominger:1994tn} where
the thunderpop was interpreted as a squeezed state in the semiclassical
theory that becomes problematic when evolved back in time. The goal
of the present section is to compute the general position of the thunderpop
$(x_{pop}^{+},x_{pop}^{-})$. 

\begin{figure}
\includegraphics[scale=0.65]{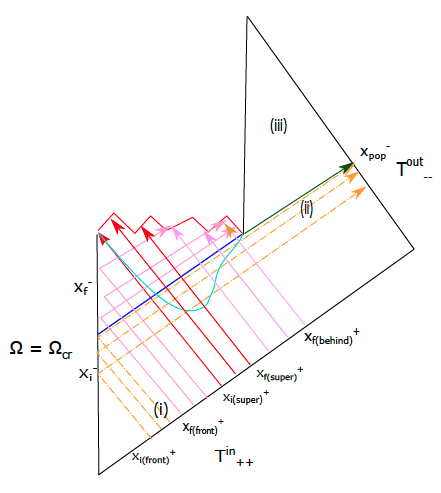}

\caption{\label{fig:General-flux-probing-1}General flux probing general black
hole. Subcritical fluxes which remain outside and pass inside the
black hole are shown in orange and purple respectively. Supercritical
flux is shown in red, the event horizon is shown in blue, and the
apparent horizon is shown in turquoise. Regions (i) ($x^{-}<x_{i}^{-}$),
(ii) ($x_{i}^{-}<x^{-}<x_{pop}^{-}$) and (iii) ($x_{(pop)}^{-}<x^{-}$)
are the natural generalizations from the RST prescription for subcritical
reflected flux. }
\end{figure}

We will take a LDV solution with additional flux sent in from some
interval in $x^{+}$ on $\mathscr{I^{-}}$ corresponding to $x_{i(front)}^{+}<x^{+}<x_{f(behind)}^{+}\leq x_{pop}^{+}$
(see figure \ref{fig:General-flux-probing-1}). We will assume the
flux turns supercritical solely in a sub-interval $x_{i(super)}^{+}<x^{+}<x_{f(super)}^{+}$
(see figure \ref{fig:General-flux-probing-1}). The consequent general
black hole solution is given by:

\begin{align*}
\Omega^{(i)}(x^{+}>x_{i(super)}^{+}) & =-\frac{x^{+}x^{-}}{\sqrt{\kappa}}-\frac{\sqrt{\kappa}}{4}\log(-x^{+}x^{-})+\Omega_{(front)}^{(i)}(x^{+})+\Omega_{(behind)}(x^{+})\\
 & +\Omega_{(super)}(x^{+}),
\end{align*}
where the total flux contribution has been separated into separate
physically significant pieces. In general, the effect of each piece
of flux is of the form:

\[
\delta\Omega(x^{+})=-\frac{x^{+}}{\sqrt{\kappa}}\delta P_{+}+\frac{\delta M_{+}}{\sqrt{\kappa}},
\]
where we are permitted to break the contributions into pieces and
linearly sum since integrals over some interval can be split and summed
over corresponding sub-intervals. 

We include a general supercritical contribution $\Omega_{(super)}$
which forms the black hole and we also separate the subcritical flux
which is reflected outside the event horizon $\Omega_{(front)}^{(i)}$
from the portion that passes behind $\Omega_{(behind)}$ and propagates
to the singularity. As the thunderpop corresponds to a null line crossing
the endpoint of black hole evaporation, it can also be taken to define
the event horizon (see figures \ref{fig:Black-hole-formation} and
\ref{fig:General-flux-probing-1}). We will assume that no flux is
sent in after the black hole has evaporated. Further, we will assume
that the flux outside the event horizon is reflected in a manner that
follows the RST prescription, hence our usage of the superscript $(i)$
on $\Omega_{(front)}^{(i)}$. 

The reflected solution $\Omega^{(ii)}$ outside the horizon is given
by applying the RST boundary conditions. In the causal past of the
thunderpop we have:

\begin{align}
\Omega^{(ii)} & =-\frac{x^{+}(x^{-}+P_{+(front)}(x^{+}))}{\sqrt{\kappa}}-\frac{\sqrt{\kappa}}{4}\log(-x^{+}x^{-})+\nonumber \\
 & \frac{M_{+(front)}(x^{+})}{\sqrt{\kappa}}+F_{(front)}(x^{-})+\Omega_{BH}(x^{+}),\label{eq:bregiontwo}
\end{align}
where we define:

\[
\Omega_{BH}\equiv\Omega_{(behind)}+\Omega_{(super)}.
\]

Evaluating $\Omega^{(ii)}$ in the region to be patched to the evaporated
LDV: $x^{+}>x_{pop}^{+}$ as well as along the thunderpop itself:
$x^{-}=x_{pop}^{-}$ results in the following simplified form for
$\Omega^{(ii)}$:

\begin{align}
\Omega^{(ii)}(x^{+}>x_{pop}^{+},x^{-}=x_{pop}^{-}) & =-\frac{x^{+}(x_{pop}^{-}+P_{+(front)}^{f})}{\sqrt{\kappa}}\nonumber \\
 & -\frac{\sqrt{\kappa}}{4}\log(-x^{+}(x_{pop}^{-}+P_{+(front)}^{f}))+\Omega_{BH}(x^{+}),\label{eq:regiontwo}
\end{align}
which can be easily patched to an appropriately shifted LDV \eqref{eq:ldv}
\begin{align*}
\Omega^{(iii)} & =-\frac{x^{+}(x_{pop}^{-}+P_{+(front)}^{f}+P_{+(behind)}^{f}+P_{+(super)}^{f})}{\sqrt{\kappa}}\\
 & -\frac{\sqrt{\kappa}}{4}\log(-x^{+}(x_{pop}^{-}+P_{+(front)}^{f}+P_{+(behind)}^{f}+P_{+(super)}^{f})).\\
\end{align*}
The continuity of such a patching entails the following form of $x_{pop}^{-}$:

\begin{equation}
x_{pop}^{-}=-P_{+(front)}^{f}+\frac{P_{+(behind)}^{f}+P_{+(super)}^{f}}{e^{-4(M_{+(super)}^{f}+M_{+(behind)}^{f})/\kappa}-1}.\label{eq:thunderpop}
\end{equation}
Using the position for the apparent horizon $\partial_{+}\Omega=0$
on \eqref{eq:regiontwo} and noting no additional flux is sent in
near the endpoint, we obtain
\begin{equation}
x_{pop}^{+}=-\frac{\kappa}{4\left(x_{pop}^{-}+P_{+(front)}^{f}+P_{+(behind)}^{f}+P_{+(super)}^{f}\right)}=\frac{\kappa\left(e^{4(M_{+(super)}^{f}+M_{+(behind)}^{f})/\kappa}-1\right)}{4\left(P_{+(super)}^{f}+P_{(behind)}^{f}\right)}.\label{eq:poptime}
\end{equation}
The determination of this general thunderpop position is the first
main technical result of this paper. 

\section{Measuring Lifetime with the Event Horizon}

The black hole region can be defined as the complement of the causal
past of $\mathscr{I^{+}}$ \citep{WaldGR}, which is bounded in part
by the event horizon. We observe that the event horizon ought to be
defined by a null line which intersects the thunderpop and is therefore
given in the region $x^{+}<x_{pop}^{+}$ by the thunderpop coordinate
$x_{EH}^{-}=x_{pop}^{-}$ \eqref{eq:thunderpop}. The endpoint of
evaporation will be given by the thunderpop as both the event horizon
and the apparent horizon meet at this point (see figures \eqref{fig:Black-hole-formation}
and \eqref{fig:General-flux-probing-1}). The initial point of black
hole formation will be the projection of the thunderpop coordinate
onto the boundary curve, $x_{BH}^{+}=\hat{x}^{+}(x_{pop}^{-})$ as
this corresponds to the edge of the event horizon. We further project
these boundary points of the event horizon onto $\mathscr{I^{-}}$,
enabling us to measure the lifetime using the asymptotically flat
light-cone coordinate $\sigma^{+}$:

\begin{equation}
x^{+}=e^{\sigma^{+}}.\label{eq:flatcoord}
\end{equation}

To determine the lifetime of the black hole, we need to determine
where the null line of the event horizon intersects the timelike boundary
curve where the RST boundary conditions are applied \eqref{eq:rstbc}.
An equation governing the boundary curve can be obtained by solving
$\partial_{+}\Omega=0$ on \eqref{eq:bregiontwo}

\[
\hat{x}^{+}=-\frac{\kappa}{4(\hat{x}^{-}+P_{+(front)}\left(\hat{x}^{+}\right))}\,.
\]
Substituting in for $x_{EH}^{-}$ gives

\begin{equation}
x_{EH}^{+}=\frac{\kappa}{4}\frac{1-e^{-4(M_{+(super)}^{f}+M_{+(behind)}^{f})/\kappa}}{P_{+(behind)}^{f}+P_{+(super)}^{f}}.\label{eq:formationtime}
\end{equation}

Converting \eqref{eq:poptime} and \eqref{eq:formationtime} to the
asymptotically flat coordinate \eqref{eq:flatcoord}, one arrives
at the following:

\[
\Delta\sigma^{+}=\log\left(\frac{e^{4(M_{+(super)}^{f}+M_{+(behind)}^{f})/\kappa}-1}{1-e^{-4(M_{+(super)}^{f}+M_{+(behind)}^{f}}}\right)=\frac{4M_{tot}}{\kappa}>\frac{4M_{+(super)}^{f}}{\kappa},
\]
where the mass of the black hole is defined as

\[
M_{tot}=M_{+(super)}^{f}+M_{+(behind)}^{f},
\]
which implies that mining does not occur. The determination of this
general lifetime is the second main technical result of this paper.
The expression is monotonically increasing in the mass and grows arbitrarily
large in the classical limit of small $\kappa.$ In particular, the
lifetime is clearly larger than in the case where probing subcritical
flux is absent. Moreover, it is manifest that this lifetime is immune
to any flux which remains outside the event horizon and in fact grows
in precisely the way one would expect for flux which passes behind
the horizon. This shows that in a model where back-reaction is treated
in a self-consistent manner, mining does not occur.

\section{Conclusions}

The issue of mining a black hole has been considered for the first
time in a quantum model with back-reaction incorporated at leading
order in a $1/N$ expansion. This provides a useful generalization
of the moving mirror model that appearing in the original paper by
Unruh and Wald \citep{UnruhWald2}, which was used to justify the
proposal that energy could be mined from Schwarzschild black holes.
We conclude that with quantum back-reaction properly incorporated,
no mining takes place in the model we consider.

The general philosophy of this kind of computation is that it correctly
captures the qualitative physics of an s-wave reduction of four-dimensional
Einstein gravity in the semiclassical limit. By modifying the model,
adding in some hopefully harmless extra terms, it can be rendered
exactly soluble. This then provides motivation to reconsider the mining
proposal of Unruh and Wald in the context of such soluble models.

A key feature of the mining proposal is that one should be able to
mine modes irrespective of their angular momentum. This is crucial
in order to get rapid mining of the type further explored in a later
essay by Unruh and Wald \citep{UnruhWald}. It would be interesting
to generalize the considerations of the present paper to this case.
A useful soluble model for this, which includes an analog of the potential
barrier for such modes is the two-dimensional charged black hole,
which has a soluble version. It is also interesting to generalize
the cosmic censorship RST boundary conditions considered in the present
work to more general energy conserving boundary conditions of the
type studied in later dilaton gravity papers \citep{ChungVerlinde,Das:1994yc,Strominger:1994xi}.

The original proposal of Unruh and Wald suggested that black holes
can be mined at a rate of a Planck energy per unit Planck time. This
would allow a rapidly mined black hole to evaporate in a light crossing
time. Such a rapid evaporation poses severe problems for extracting
quantum information from black holes, in holographic approaches to
quantum gravity. Subsequent work incorporating classical back-reaction
led to the minimum mining lifetime being walked back to a time of
$O(M^{2})$ \citep{Brown:2012un}. The conclusions of the present
work support the idea that mining of Schwarzschild black holes cannot
occur once quantum back-reaction is incorporated, potentially strengthening
the conclusions of Brown.

\bibliographystyle{utphys}
\bibliography{RSTMiningBib}

\end{document}